# A Route to High-Toughness Battery Electrodes

*Nicola Boaretto[a,b], Jesús Almenara[c], Anastasiia Mikhalchan[a], Rebeca Marcilla[b,*], Juan J. Vilatela[a,*]*

[a]IMDEA Materials, Eric Kandel 2, 28906 Getafe, Madrid, Spaintable

[b]IMDEA Energy, Avda. Ramón de la Sagra 3, 28935 Móstoles, Madrid, Spain

[c]Escuela Politécnica Superior, Universidad Carlos III, Avda. de la Universidad 30, 28911 Leganés, Madrid, Spain

*Correspondence to: RM rebeca.marcilla@imdea.org, JJV juanjose.vilatela@imdea.org

ABSTRACT. There is an increasing interest in materials that combine energy storing functions with augmented mechanical properties, ranging from flexibility in bending to stretchability to structural properties. In the case of lithium ion batteries (LIBs), these mechanical functions could enable their integration in emerging technologies such as wearable, free-form electronics and ultimately as structural elements, as example in transport applications. This work presents a method to produce flexible LiFePO$_4$ (LFP) electrodes with an extraordinary combination of electrochemical and mechanical performance. Such electrodes exhibit an exceptionally high specific toughness of 1.6 J g$^{-1}$, combined with superior rate capability (29 % increase of the specific capacity at 500 mA g$^{-1}$, even with 60 % reduced conductive additive content) and energy density (60 % increase at 500 mA g$^{-1}$, on a LFP/Li full cell basis), with respect to reference electrodes with typical metallic current collectors. These properties are a result of the strong adhesion of the active material particles to the high surface area carbon nanotube fiber (CNTF) fabric, used as lightweight, tough and highly conducting current collector. This strong adherence minimizes electrical resistance, mitigates interfacial failure and increases ductility through heterogeneous strain after cohesive failure of the inorganic phase. As a result, these electrodes can withstand large deformations before fracture (above 15 % tensile deformation) and, even after fracture they retain excellent electrochemical performance, with a full-electrode-normalized specific capacity of 90 mAh g$^{-1}$ at 500 mA g$^{-1}$, approximately double that of unstretched, Al-supported, LFP electrodes with equivalent loading.

1. INTRODUCTION

Electrochemical energy storage devices, and lithium-ion batteries (LiBs) in particular, play a fundamental role in the ongoing energy transition from fossil fuels to renewable energy sources.[1] The fields of application of LiBs range from consumer electronics, to the automotive and to the stationary energy storage sectors. In addition, new applications are emerging, such as wearable batteries[2–4] or even multifunctional batteries that can act as structural elements (i.e. structural batteries).[5,6]

The research on standard lithium-ion batteries focuses mostly on improving the safety and energy density,[7–9] but the development of mechanically-augmented batteries includes optimizing additional mechanical parameters alongside.[10] Several flexible battery configurations have been developed, such as fiber or cable batteries,[11–19] patterned array configurations,[20] kirigami/origami configurations,[21,22] batteries with electrodes supported on flexible substrates,[23] and accordion-like stretchable batteries.[24]

The most attractive flexible cell configurations involves the use of carbonaceous materials as current collectors, including carbon fibers,[25,26] graphene,[27,28] or carbon nanotubes (CNT).[29–39] Assembled as networks of conductive elements, carbon based current collectors not only withstand large strains, but can also increase energy density by reducing up to 15 % of the total cell mass relative to metal current collectors.[5] CNT-based films are particularly well suited for this purpose, due to their lightweight, high electrical conductivity at small volume fractions and mechanical robustness. CNT-based current collectors have been successfully applied at the cathode, in particular with $LiFePO_4$ (LFP)[32,40,41] and $LiCoO_2$ (LCO),[17,29,30,37–39,42] and at the anode, mostly with $Li_4Ti_5O_{12}$ (LTO).[17,30,32,36,38,39,41,43]

Because of the large deformation involved, producing materials/architectures for stretchable battery electrodes is far more challenging than developing foldable electrodes. CNT-based electrodes have been central to the development of bendable and stretchable batteries, showing excellent retention of the electrochemical performances even under repeated mechanical deformation. However, most architectures explored so far rely on depositing CNTs onto pre-strained elastomers,[36,43] forming co-axial/twisted microfilaments,[15,17,44] or other geometric dispositions that circumvent the challenge of simultaneously enabling high strains and high stresses. The tensile strength and strain at break of CNT-supported electrodes are usually between 1 MPa and 7 MPa and below 5 %, respectively,[31,37,39,41] with high strength severely compromising ductility.[37,39] Indeed, we find that literature values for toughness (fracture energy) are in the range 0.01 – 0.1 J g$^{-1}$, similar to a brittle polymer.

Producing electrodes with higher toughness, i.e. that can withstand substantial stresses as well as strains, would not only increase safety by avoiding overstretching, but would also enable the integration of tough battery electrodes in/as structural composites.

A significant enhancement of the electrodes mechanical properties could be achieved by using as current collectors CNT fiber (CNTF) fabrics, which are characterized by exceptionally high electrical conductivity, porosity and toughness.[45] As an example, CNTF have been used already as electrodes in double layer capacitors,[46,47] and in the fabrication of fiber- and cable-shaped electrochemical storage devices for wearable applications.[48]

In this report, we explore the fabrication and the mechanical, morphological and electrochemical properties of large-area planar LFP electrodes with a continuous network of CNTF playing the dual role of as scaffold and current collector (CNTF-LFP). The electrochemical performance of the resulting composite electrodes is above that of Al-LFP electrodes, with tensile strength about 15 MPa, ductility as high as 15 % and, very importantly, a specific toughness of 1.6 J g$^{-1}$, one order of magnitude above that of related electrodes.[27,31,37,39,41,49–51] Electrodes subjected to uniaxial tests up to fracture and subsequently tested in half-cell configurations show nearly identical electrochemical properties as unstretched ones. Key to realize all these properties is the exceptional combination of high tensile fracture energy and high electrical conductivity, and large porosity of the constituent fabric of CNT fibers.

2. EXPERIMENTAL SECTION

Carbon nanotube (CNT) fiber fabrics were synthesized by direct spinning of CNTF from the gas phase via floating catalyst CVD,[52] using conditions to produce electrodes reported previously,[46] Synthetic conditions were adjusted to obtain thin CNTF fabrics of ~5 μm thickness, with density of about 0.3 g cm$^{-3}$.

LiFePO$_4$ (LFP) electrodes were prepared by mixing the active material (Aleees, LFP-NCO M121) with polyvinylidene fluoride (PVDF, MTI, MW 600k) as binder and carbon black (Imerys, Super C65) as conductive additive in mass proportion 90:5:5. LFP and the conductive additive were pre-mixed in a planetary ball mill (Retsch, PM 100) at 300 rpm for 30 min. After drying for two hours at 110 °C, the LFP/C65 powder was dispersed in a 5 wt% solution of PVDF in 1-methyl-2-pyrrolidone (NMP, purity ≥ 99%, Sigma Aldrich), with a disperser (IKA ULTRA-TURRAX, T 10). The slurry was coated by doctor blade with a semi-automatic film applicator (COATMASTER 510, Erichsen) and a micrometer-adjustable film casting knife (MTI, EQ-Se-KTQ-100) using a blade heights of 100 μm, directly on CNTF fabrics. The CNTF-LFP electrodes were finally dried overnight at 80 °C, under vacuum. The resulting average thicknesses were of 35 μm, corresponding to a total loading of 3.8 mg cm$^{-2}$ and to an active material loading of 3.3 mg cm$^{-2}$. LFP electrodes on aluminium (MTI, EQ-bcaf-15u-280, purity > 99.3 %, 15 μm), with mass ratio LFP:SC65:PVDF of 80:15:5, were also prepared (Al-LFP). In this case, the average thickness was of 42 μm and the total loading of 4.6 mg cm$^{-2}$, corresponding to a LFP loading of about 3.6 mg cm$^{-2}$. The electrochemical properties of Al-LFP electrodes were compared with those of commercial LFP electrodes (1 mAh cm$^{-2}$, 83 % active material).

Uniaxial tensile tests on coated CNTF fabrics were performed with a displacement-control dual column table top universal testing system (INSTRON 5966, load cell 500 N) at a constant strain rate of 1 mm min$^{-1}$. Coin cells (CR2032, TOB) with half-cell configuration were assembled with LFP electrodes as cathode, lithium foil as anode (Sigma Aldrich, 400 μm thickness), 1 M solution of LiPF$_6$ in 1:1 vol % ethylene carbonate/diethyl carbonate as electrolyte (300 μL, Solvionic, 99.9 %, H$_2$O < 20 ppm), and Whatman GF/D glass fiber as separator. LFP electrodes were dried overnight at 80 °C under vacuum before assembling, and cell assembly was carried out in a glove box, under Argon atmosphere. Electrochemical tests were performed with a Neware BTS 4000 battery tester. At first, a rate capability test was carried out: the cells were cycled with CCCV protocol between 2.5 V and 3.8 V, at the following current densities: 17, 35, 80, 165, 500, 17 mA g$^{-1}$ (referred to the mass of LFP). In the following aging test, the cells were cycled 100 times at 80 mA g$^{-1}$. Electrochemical tests were performed also on the samples after mechanical tests up to fracture. Scanning electron microscopy (SEM) images were obtained using FIB-FEGSEM dual-beam microscope (Helios Nanolab 600i, FEI) at 5 kV and 0.69 nA.

Impedance spectra on pristine and cycled Al-LFP/Li and CNTF-Li coin cells were acquired with a BioLogic VMP3 potentiostat, in the frequency range between 1 MHz and 10 mHz and with a signal amplitude of 10 mV. Impedance spectra were acquired also on Swagelok T-cells with CNTF-LFP and Al-LFP as working electrodes, and Li as both counter and reference electrode.

## 3. RESULTS

### 3.1. MORPHOLOGY AND MECHANICAL PROPERTIES

CNTF-LFP electrodes were fabricated through the standard process of casting a slurry of active material onto CNTF, that is acting as metal-free current collector, without having to modify the deposition process (Figure S1a,b,c). The coating resulted in homogeneous CNTF-LFP electrodes with thickness between 35 and 60 μm (CNTF included), no cracking, and good flexibility. The back side of CNTF current collector is shiny due to polymer infiltration all through the CNTF fabric thickness (Figure S1d). The final mass fraction of active material in the electrode assembly was of 82 % for CNTF-LFP electrodes, whereas it was of about 42 % for Al-LFP electrodes.

Inspection of the electrode microstructure (CNTF-LFP) using scanning electron microscopy (Figure 1) shows that active material is firmly attached to the thin (5 μm) CNTF fabric (Figure 1a,b). In fact, optical inspection and electron micrographs confirm that the binder, the carbon black particles (with particle size of about 50 nm) and the smallest active material particles infiltrate through the porous fabric down to the back side of the CNTF current collector that was not originally exposed to the slurry during coating (Figure 1b,d). Capillary forces drive infiltration of the liquid slurry and favor the flow of small particles into the current collector pores. The resulting structure can be visualized as a composite comprised of an electrode with a built-in porous current collector. The remaining active material forms a "coating" that preserves its expected morphology, with irregular-shaped 1-2 µm LFP particles and smaller carbon black particles bonded by PVDF (Figure 1c).

In-house Al-LFP electrodes (Figure S2) have thickness of about 50 µm, with current collector thickness of 19 µm (Figure S2a). The SEM images show a clear mismatch between the coating and the current collector (Figure S2a and S2b). This is expected to affect negatively the electrode performance. The active material particles have obviously similar shape and size to CNTF-LFP electrodes (Figure S2c and S2d). With respect to the CNTF-LFP electrodes, it is possible to appreciate the higher concentration of carbon black, which is necessary to compensate the worse contact with the current collector.

The coating in commercial Al-LFP electrodes (Figure S3) has thickness of about 50 µm, and the current collector has about 20 µm thickness (Fig. S3a). Between the current collector and the active material coating, an interlayer of 2 – 3 µm thickness is observed, mainly composed of conductive additive and binder (Fig. S3b). The probable function of this interlayer is to enhance the charge transport between current collector and electroactive layer. The active material has particle size < 1 µm and the particle shape is also more regular than the one used in the in-house electrodes (Figure S3c and S3d). Besides the active material and the carbon black, other prismatic particles with size of about 2 µm are observed, which suggest the presence of conductive graphite as additive.

In order to investigate the potential application of CNTF as mechanical stable current collector in flexible/stretchable devices, the longitudinal mechanical properties of CNTF-LFP electrodes were determined by uniaxial tensile testing and compared with Al-LFP electrodes. Figure 2 shows that Al-LFP electrodes exhibit essentially the stress-strain curve of the aluminium current collector, since the coating carries negligible load and is very brittle. The electrodes average tensile strength is at around 50 MPa, 26 MPa cm$^3$ g$^{-1}$ in specific units (Table 1), which corresponds to a tensile strength of 140 MPa for the bare Al current collector, close to the value reported by the manufacturer (≈ 150 MPa). Average elongation at break is only 1 %, leading to low toughness of 0.3 J cm$^{-3}$ or 0.2 J g$^{-1}$ in specific units. Similar values were observed also for commercial LFP electrodes with areal capacity of 1 mAh cm$^{-2}$ (see Figure S4).

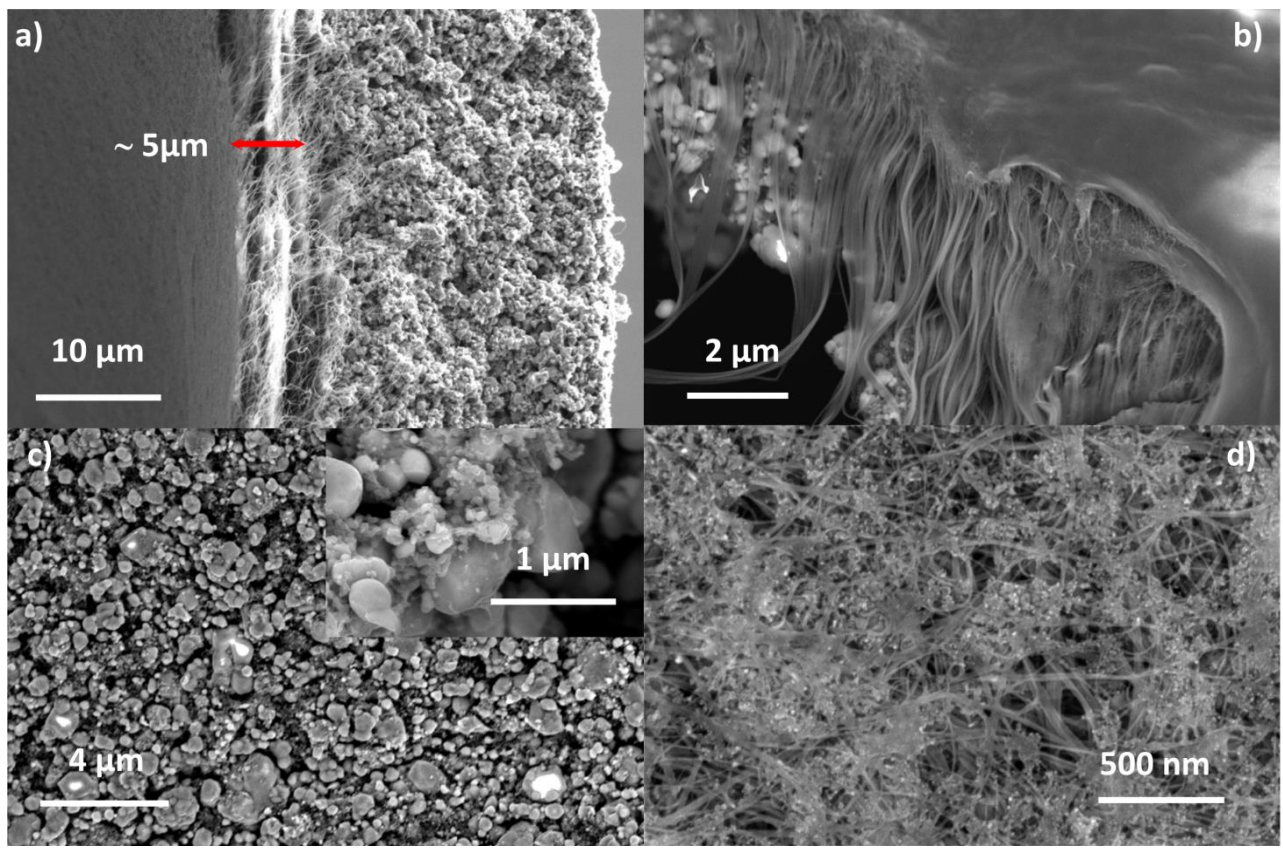

**Figure 1.** SEM images of a CNTF-LFP electrode: a) side view; b) side/back view; c) top view (LFP side, higher magnification in the inset); d) bottom view (current collector side).

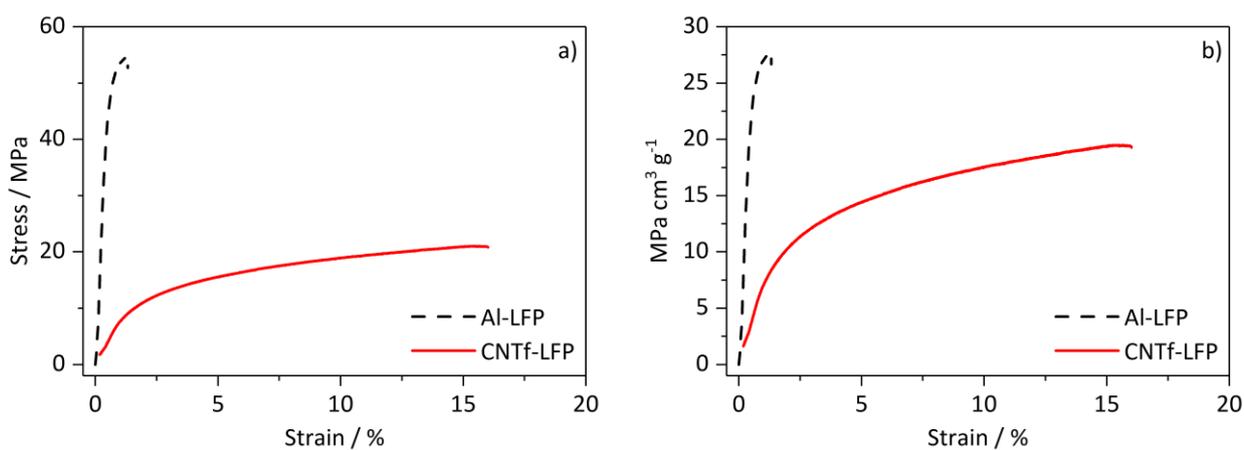

**Figure 2.** a) Stress-strain curves of Al-LFP and CNTF-LFP electrodes; b) Specific stress versus strain.

**Table 1.** Mechanical properties of CNTF-LFP and Al-LFP electrodes (average of three samples).

|  | Specific tensile strength / MPa cm$^{-3}$ g$^{-1}$ | Specific Modulus / GPa cm$^3$ g$^{-1}$ | Strain at break / % | Specific toughness / J g$^{-1}$ |
|---|---|---|---|---|
| Al-LFP | 26 ± 5 | 6.8 ± 0.9 | 0.9 ± 0.4 | 0.2 ± 0.1 |
| CNTF-LFP | 15 ± 5 | 0.9 ± 0.3 | 13 ± 6 | 1.6 ± 0.9 |

The CNTF-LFP electrodes have slightly lower specific strength 15 MPa g$^{-1}$ cm$^3$, which is attributed to the much lower current collector mass ratio ($m_{cc} < 0.1$). But very interestingly, their average strain-to-break is > 10 %, reaching in the best cases over 15 % (Figure 2). The stress strain curves for CNTF-LFP electrodes show an elasto-plastic behavior, with progressive yielding above 1 % deformation. Because of the large deformation sustained by these electrodes, the average specific toughness reaches 1.6 J g$^{-1}$ (Table 1), with a maximum of 2.4 J g$^{-1}$ observed. This toughness is one order of magnitude higher than that of Al-LFP electrodes and between one and two orders of magnitude over previous reports on flexible planar electrodes with nanocarbon-based current collectors (see Table 2). For reference, it is above a structural thermosetting resin or human bone, although lower than that of carbon fiber structural electrodes (4.3 J g$^{-1}$).[6]

**Table 2.** Comparison of combined electrochemical and mechanical properties of flexible electrodes in planar configuration. Details on the calculations are reported in the Supplementary Information.

| Electrode type | Toughness | | Active mass fraction | Sp. Capacity | | Ref. |
|---|---|---|---|---|---|---|
|  | J cm$^{-3}$ | J g$^{-1}$ |  | mAh g$^{-1}$ | mAh cm$^{-2}$ |  |
| CF/SBE/Cu | 5 | 4.3 | 0.07 | 16 | 0.33 | 6 |
| CNTF-LFP | 1.7 | 1.6 | 0.81 | 132 | 0.5 | This work |
| LTO-AgNW$MF | ~0.5 | ~0.2 | 0.7 | 110 | 0.8 - 3 | 23 |
| Si/CNT | 0.25 | 0.19 | 0.47 | 494 | 1.16 | 49 |
| LFP/CNT | 0.022 | 0.04 | 0.95 | 150 | 0.073 | 41 |
| LTO/CNT | 0.021 | 0.04 | 0.95 | 151 | 0.075 | 41 |
| LTO/rGO* | 0.09 | 0.045 | < 0.16 | 26 | 1.3 | 27 |
| LCO/rGO* | 0.09 | 0.045 | < 0.16 | 26 | 1.3 | 27 |
| Si/MX | 0.007 | 0.014 | 0.09 | 288 | 2.9 | 50 |
| LCO/PA/CNT | 0.01 | 0.01 | 0.57 | 76 | 1.2 | 39 |
| LTO/PA/CNT | 0.01 | 0.01 | 0.59 | 78 | 1.34 | 39 |
| Graphite/SACNT | 0.006 | 0.007 | 0.8 | 268 | 1.92 | 31 |
| LCO/SACNT | 0.004 | 0.002 | 0.97 | 146 | 0.6 | 37 |
| LFP/C | 0.003 | (~0.003) | 0.8 | 125 | 0.3 | 51 |

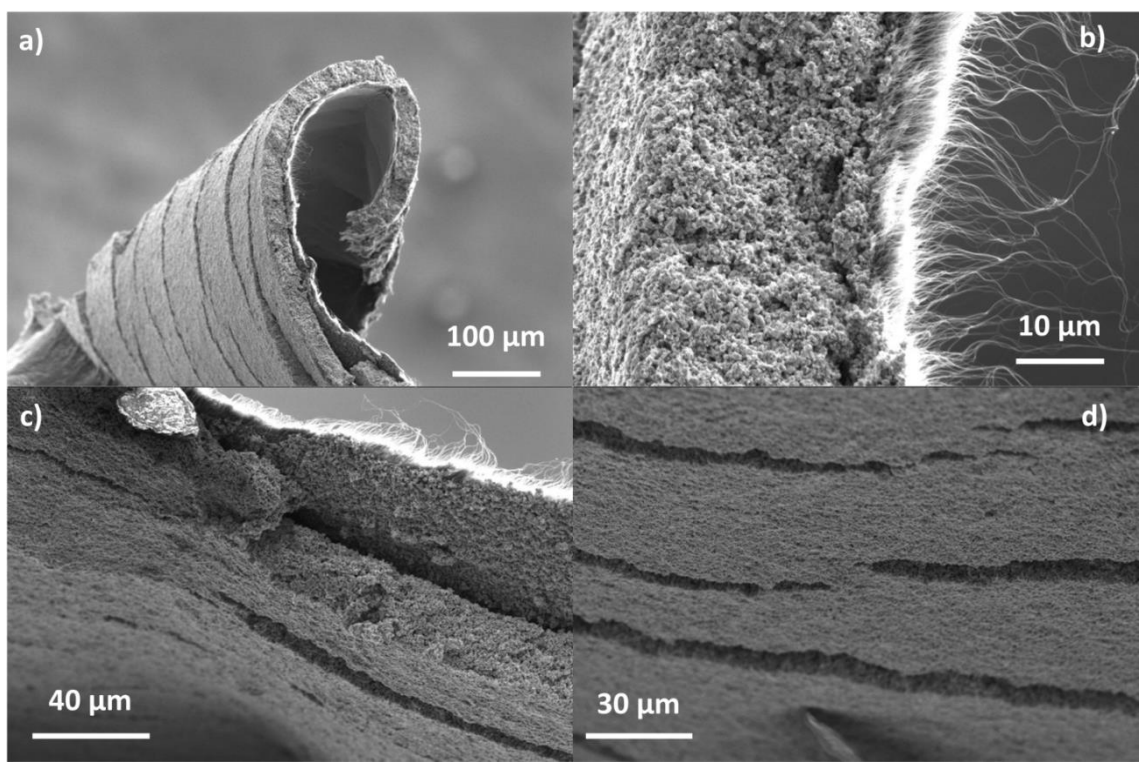

**Figure 3.** Cross-sectional SEM images of a stretched CNTF-LFP electrode: a) overview of the folded CNTF-LFP electrode at low magnification (200×); b) side view (2000×); c) image of the coating-current collector interface with partly detached coating layer; d) tilted view of the coating surface.

The high ductility and toughness of the CNTF-LFP electrodes contrasts with the expected brittleness of the active material. Fractography analysis by electron microscopy on electrodes stretched to fracture shows key features enabling these properties. The most striking feature is the strong adhesion of LFP to the CNTF fabric; the coating remains attached to the current collector even after fracture (Figure 3a,b). There is clear evidence of cracks in the LFP coating, running perpendicular to the direction of mechanical loading. They develop when the strength of the coating is exceeded (around 0.2 MPa, see Figure S5) at the beginning of the test, but because of the strong interfacial adhesion with the CNTF fabric, these regions remain firmly attached and the mechanical integrity of the full electrode is preserved. Once the cracks develop, the local region in the underlayer of CNTF fabric can undergo further longitudinal deformation. At that point, the strain field is no longer uniform in the electrode and the majority of the deformation proceeds in the CNT fabric, albeit without causing apparent damage to the LFP-rich regions.

The regular spacing (~ 100 μm, Figure 3c,d) between cracks in the LFP coating is indication of the stress transfer length. It corresponds to the length along the coating over which stress is transferred from the CNTF by shear. To a first approximation, the following relation holds:

$$l = \sigma_c \frac{t}{\tau} \quad (1)$$

Where $\sigma_c$ is the tensile strength of the coating, $l$ is the stress transfer length, $\tau$ is the shear stress at the interface and $t$ the coating thickness. Since $l > t$, the shear stress is lower than the tensile strength coating. We measured a tensile strength of about 0.2 MPa for an LFP film (see Figure S5), which is close to the value of 0.25 MPa reported in [41]. It can be therefore concluded that $\tau < 0.1$ MPa.

In fact, inspection of the fracture surface, as in the example in Figure 3c, shows local damage at the interface between bulk LFP and the LFP/CNTF layer, an indication that the latter has distinct properties and is stronger than the bulk LFP coating.

Overall, this analysis highlights the fact that the crack spacing, and thus the number of cracks in the electrode, could be reduced in the future by increasing the strength of the coating, for instance through the use of emerging binders and fillers.[53]

In summary, CNTF-LFP electrodes are characterized by combined high tensile strength and ductility, resulting in order-of-magnitude higher fracture energy than planar CNT-supported electrodes.[31,37,39,41,49] The interest then, is in studying the electrochemical properties of these electrodes before and after fracture.

## 3.2. ELECTROCHEMICAL PROPERTIES

CNTF-LFP and Al-LFP electrodes were tested in half cell configuration by galvanostatic cycling at different current densities (17 – 500 mA g$^{-1}$, on the LFP mass).

Figures 4a and 4b show the evolution of the potential profiles of two representative cells with CNTF-LFP electrodes Al-LFP electrode, respectively. At low current densities, both batches of electrodes show the typical straight plateau at about 3.4 V vs Li/Li$^+$, and a steep potential drop/increase towards the end of discharge/charge. However, at high current densities Al-LFP electrodes have significantly higher polarization, with steeper voltage slope and higher average voltage difference between the charge and discharge plateaus. The latter, at 500 mA g$^{-1}$, is 0.22 V for CNTF-LFP electrodes and almost 0.5 V for Al-LFP electrodes. This translates into a higher discharge capacity for CNTF-LFP electrodes for current densities $J >$ 80 mA g$^{-1}$. At 500 mA g$^{-1}$, the average specific discharge capacity (CC discharge till 2.8 V) is 124 mAh g$^{-1}$ for CNTF-LFP and 96 mAh g$^{-1}$ for Al-LFP (Figure 4c). It must be highlighted that LFP electrodes deposited onto CNTF perform much better than those deposited onto Al, even though the latter have much higher carbon black content (15 % versus 5 % of CNTF-LFP electrodes), and that the rate capability of in-house Al-LFP electrodes is comparable to that of commercial LFP electrodes (Figure S6).

These results indicate that the impregnation of active material into the CNTF current collector, besides being beneficial in terms of mechanical properties, also enhances the electrical contact between active material and current collector. It must be also stressed, that attempts to reduce the concentration of carbon black in Al-LFP electrodes resulted in even higher polarization and poor electrochemical performances.

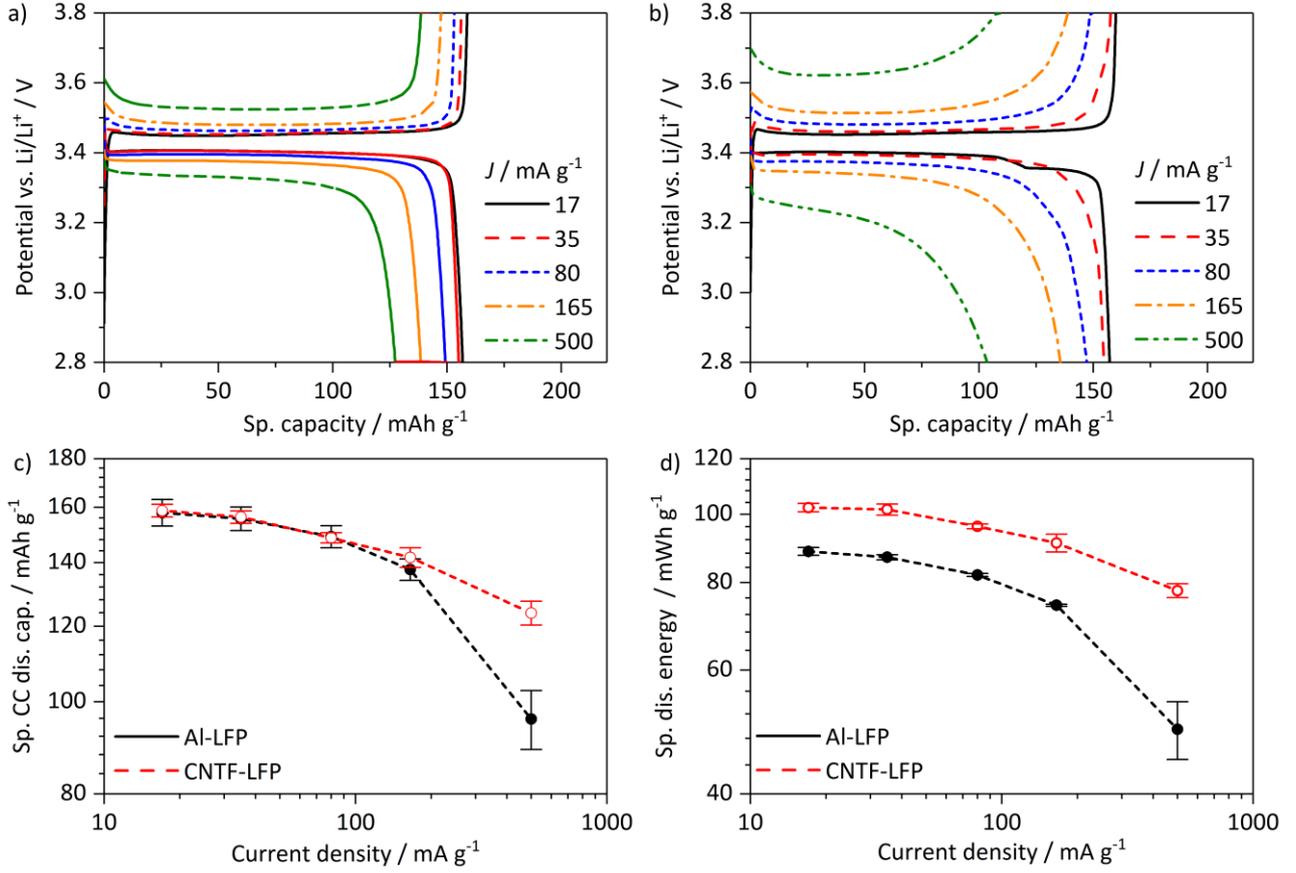

**Figure 4.** Potential profiles of LFP/Li cells during rate capability test with: a) CNTF-LFP/Li cell; b) Al-LFP/Li cell; c) specific discharge capacity (CC discharge till 2.8 V, normalized on the LFP mass) as a function of the current density (normalized on the LFP mass). d) Specific discharge energy (normalized on the total cell mass) on the current density (normalized on the LFP mass).

Recently, Tian et al. proposed a novel method to quantify the rate performances of battery electrodes.[54] The model considers the (specific) capacity $Q$ as a function of the fractional rate $R$, which is the ratio between the current (density) and the experimental (specific) capacity at the correspondent current value. The dependence is described by the following empirical equation:

$$Q = Q_0[1 - (R\tau)^n(1 - e^{-(R\tau)^{-n}})] \qquad (2)$$

where $Q_0$ is the maximum capacity at low rates, $\tau$ is a time constant describing the transition between low rate and high rate regimes, and $n$ should be, in theory, comprised between 0.5 and 1. For diffusion-limited behavior, $n = 0.5$, and for resistance-limited behavior $n = 1$. The time constant $\tau$ is used to calculate the transport coefficient $\Theta = L^2/\tau$, which takes into account the electrode coating thickness $L$. The transport coefficient is then used to compare the rate performances of the electrodes.

The model was applied to compare the performance of CNTF-LFP and Al-LFP electrodes. The rate tests performed in this study cover the low rate regime and the initial part of the transition region. In this range, $R\tau \ll 1$, and the contribution of the exponential factor is negligible. It is possible therefore to simplify equation 1 as follows:

$$Q = Q_0[1 - (R\tau)^n] \qquad \text{Eq. 2}$$

The fitting parameters are reported in Table 3, whereas experimental and fitting curves are shown in the supplementary material (Figure S7). CNTF-LFP electrodes have $\tau$ values close to 45 s, whereas Al-LFP electrodes have $\tau$ close to 180 s. In other words, the transition to the high-rate regime occurs at higher rates for CNTF-LFP electrode, with a characteristic rate $R_c = 1/\tau$ of ca. 80 h$^{-1}$ for CNTF-LFP electrodes and 20 h$^{-1}$ for Al-LFP electrodes. As noted above, in order to compare the rate performances, also the electrode thickness needs to be considered. CNTF-LFP electrodes have transport coefficients of ca. $2 \cdot 10^{-11}$ m$^2$ s$^{-1}$, versus $4 \cdot 10^{-12}$ m$^2$ s$^{-1}$ of Al-LFP electrodes, thus confirming that the former have better rate performance (commercial electrodes have intermediate values of $1 \cdot 10^{-11}$ m$^2$ s$^{-1}$, see Table S1). The obtained values fall in the expected range for lithium battery electrodes.[54] In addition, we obtained values of $n$ of 0.4 for CNTF-LFP electrodes, suggesting a diffusion-limited behavior. On the contrary, both in-house and commercial Al-LFP electrodes have $n$ values close to 0.7, indicating mixed diffusion-resistive limitation. In general, higher values of $n$ correspond to faster capacity decrease at high rates. This further confirms that the use of CNTF as current collector reduces the resistance limitations, thus enhancing the electrode rate capability.

**Table 3.** Fitting parameters obtained for rate curves of CNTf-LFP and Al-LFP electrodes.

|  | $L$ / μm | $Q_o$ / mAh | $\tau$ / s | $n$ | $\Theta$ / cm$^2$ s$^{-1}$ |
|---|---|---|---|---|---|
| CNTF-LFP | 33 ± 10 | 168 ± 7 | 45 ± 25 | 0.4 ± 0.1 | $(2 \pm 3)\cdot 10^{-7}$ |
| Al-LFP | 27 ± 5 | 163 ± 1 | 176 ± 11 | 0.67 ± 0.04 | $(4 \pm 2)\cdot 10^{-8}$ |

The specific energy of the tested cells (Figure 4d) was calculated by considering the total cathode mass (including current collector), a lithium loading of 1 mg cm$^{-2}$, and electrolyte loading of 12 mg cm$^{-2}$ (which implies a large excess of lithium and electrolyte loading close to a commercial lithium-ion battery[5]). At 17 mA g$^{-1}$, cells with CNTF-LFP electrodes deliver about 102 mWh g$^{-1}$, versus 89 mWh g$^{-1}$ of the cells with Al-LFP electrodes. At higher current densities, the difference is even more marked (79 mWh g$^{-1}$ versus 49 mWh g$^{-1}$ at 500 mA g$^{-1}$), due to the higher specific capacity obtained with CNTF-LFP electrodes. This corresponds to an energy density increase of the 15 % and 60 %, respectively. Up to 80 mA g$^{-1}$, the energy gain is attributed to the higher fraction of active material of CNTF-LFP electrodes. Indeed, if the energy is normalized on the mass of active material, similar values are obtained (ca. 550 mWh g$^{-1}$ at 17 mA g$^{-1}$). The higher active material fraction results from: a) the small contribution from the CNTF current collector, and b) the lower required amount of carbon black. The ratio of active material with respect to the total cathode mass including current collector is above 0.8 for CNTF-LFP electrodes, compared with 0.4 for Al-LFP electrodes, respectively. Higher energy densities can be obviously achieved by increasing the mass loading. Indeed, cells with CNTF-LFP electrodes, with an active material loading of ca. 5 mg cm$^{-2}$, delivered energy densities of 153 mWh g$^{-1}$ and 127 mWh g$^{-1}$, at 17 and 500 mA g$^{-1}$, respectively (Figure S8). These values are higher than the ones obtained with 7 mg cm$^{-2}$ commercial Al-LFP electrodes, which delivered 130 and 98 mWh g$^{-1}$, in the same conditions. The energy density, however, depend on the cell configuration. As the cathode mass loading increases, the relative contribution of the current collector decreases, therefore the gain due to the utilization of CNTF current collectors decreases as well. On the other end, the energy density gain due to the lower amount of carbon black is, in theory, independent on the cathode mass loading. Finally, the energy gain depends on the mass fraction of the cathode in the whole cell. If the mass contribution of the other cell components increases, as example by using graphite instead of Li metal at the anode, the energy density gain is bound to decrease. In commercial Li-ion cells, the mass contribution of the current collectors is between 10 and 15 %[5] which can be set as theoretical limit for the energy density gain due directly the elimination of the metallic current collectors (without considering any decrease of the conductive additive content and any potential performance enhancement).

In any case, the results demonstrate that it is possible to achieve an increase of the energy density by using CNTF current collectors, due to three factors: a) decrease of the current collector weight; b) enhanced contact between active material and current collector, enabling a reduction of the conductive additive content; c) higher specific capacity at high current density.

After the rate capability test, the cells were charged and discharged between 3.0 V and 3.8 V, for 100 cycles, at about 80 mA g$^{-1}$. Figures 5a and 5b show the evolution of the potential profiles of a CNTF-LFP cell and of an Al-LFP/Li cell, respectively. The initial capacity is in both cases about 145 mAh g$^{-1}$, although cells with CNTF-LFP electrodes show significantly flatter potential profiles and lower IR-drop. Upon cycling, cells with Al-LFP electrodes show marked polarization increase, which results in an average capacity retention of 85 % (CC discharge till 3.0 V) after 100 cycles (best performing cell in Figure 5c: 92 % after 100 cycles). Commercial electrodes showed even lower capacity retention (74 %, Figure S9), probably due to the higher loading. On the contrary, only marginal polarization increase is observed with CNTF-LFP electrodes, corresponding to an average 96 % CC capacity retention after 100 cycles (best performing cell in Figure 5c: 99 % after 100 cycles). Correspondently, the coulombic efficiency is also higher with CNTF-LFP electrodes, decreasing from about 99.8 % to 99.0 % throughout the test (Figure 5c). In comparison, the coulombic efficiency of cells with Al-LFP electrodes decreases from 99.1 % to 98.6 % after 100 cycles. The observed fluctuations in the capacity and efficiency of CNTF-LFP/Li cells (Figure 5c) were observed on different cells, which were being cycled contemporarily, at the same absolute time (see Figure S10). Thus, they are attributed to instrumental factors, such as temperature fluctuations.

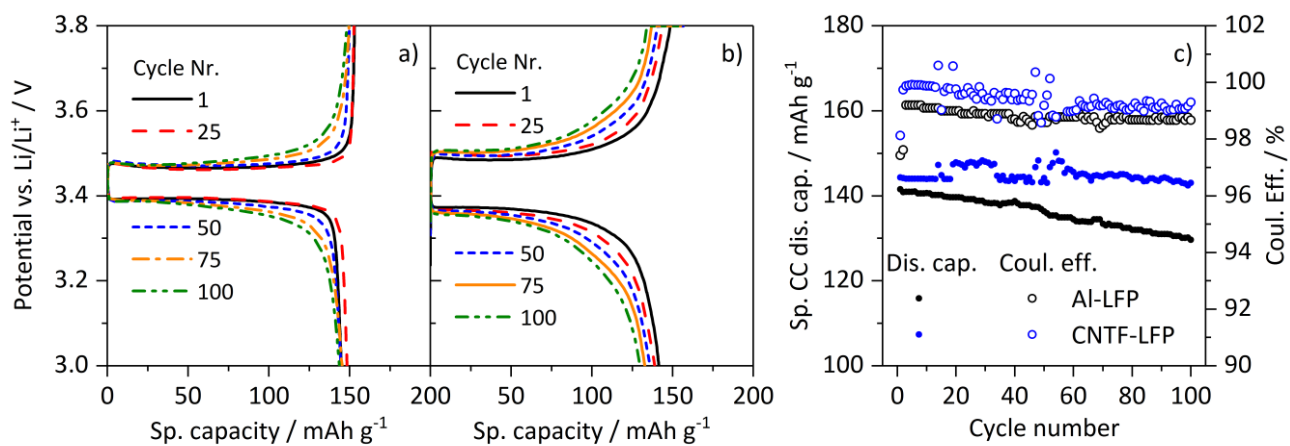

**Figure 5.** Evolution of the potential profiles during cycling experiment of: a) a CNTF-LFP/Li cell; b) a Al-LFP/Li cell; c) Specific CC discharge capacity at 3.0 V and coulombic efficiency of a CNTF-LFP/Li and an Al-LFP/Li cell.

The higher capacity retention and coulombic efficiency of CNTF-LFP electrodes suggest that the latter are superior also in terms of cyclability, probably because the formation of a composite interface between active material and current collector is able to maintain a good electrical contact even upon prolonged cycling.

Impedance spectroscopy performed on pristine and cycled cells shows slightly lower resistance for CNTF-electrodes, both in pristine and cycled cells (Figures S11a and FigS11b, respectively). However, no definitive conclusion can be obtained regarding the cathode contribution alone, as these measurements were performed on two-electrode cells, and therefore the spectra contain also the contribution of the lithium anode. Indeed, the cell resistance decrease observed after cell cycling is attributed to the roughening of the lithium surface. If the spectra are acquired on thee-electrode cells (thus with the possibility of separating the contribution of working and counter electrode), it is possible to observe that the main contribution to the middle frequency semi-circle is almost entirely due to the counter electrode, i.e. to the charge transfer/SEI resistance at the lithium metal anode (Fig. S11c). After subtracting the counter electrode contribution, no obvious difference is observed between the spectra of Al-LFP and CNTF-LFP electrodes. What is certain is that Al-LFP electrodes require a much higher amount of conductive additive than CNTF-LFP electrodes, as it appears from Fig. S11d. Here, the impedance spectra of Al-LFP/Li cells with 5 %, 10 % and 15 % conductive additive are represented. Clearly, effective charge transfer between active material and current collector is achieved only with 15 % conductive additive, whereas CNTF-LFP electrodes have similar resistance with only 5 % of carbon black.

Electrochemical cycling tests were carried out on CNTF-LFP electrodes strained to fracture, in order to analyze the effects of the stretching on rate capability and cyclability. It must be noted that the tensile strain to which the electrodes have been subjected is higher than the one which would result from typical bending tests. In a bending test, the surface strain is approximately $\varepsilon = t/R$, where $t$ is the thickness of the electrode, measured from the bending mandrel surface, and $R$ is the curvature radius. For a curvature radius $R$ of 1 mm, which is usually the smallest mandrel radius in a bending test, and with an electrode thickness of 30 µm, the maximum strain is thus 0.03, much smaller than the strain applied in the tensile test (average strain at break 0.13). Reversely, the bending radius necessary to obtain a surface strain of 0.13 by bending is equal to 0.23 mm, which is equivalent to folding the electrode.

Similarly, in case of a bending test on a full-cell, two limit cases can be distinguished, 1) all cell layers (electrodes and separator) are cohesive and 2) the cell layers are independent. In the first case, the strain at the outermost layer will be proportional to the cell thickness. If the cell thickness is 100 µm (without considering the casing, which is regarded as independent), the maximum tensile strain, on the outermost current collector, is equal to 0.1. In the second case, each layer can be considered as independent, experiencing a strain which is proportional to its own thickness, i.e.

between 0.03 and 0.05. These considerations highlight the much more severe mechanical conditions undergone by the electrodes tested after tensile fracture compared to regular cyclic bending tests.

In the rate capability test, the potential profiles resemble the ones of pristine electrodes (Figure 6a). Steeper potential plateau slope is observed only at 500 mA g$^{-1}$. Correspondingly, the CC discharge capacity is similar to the one obtained with pristine electrodes. Only at 500 mA g$^{-1}$, lower average CC discharge capacity of 112 mAh g$^{-1}$ is observed (Figure 6b).

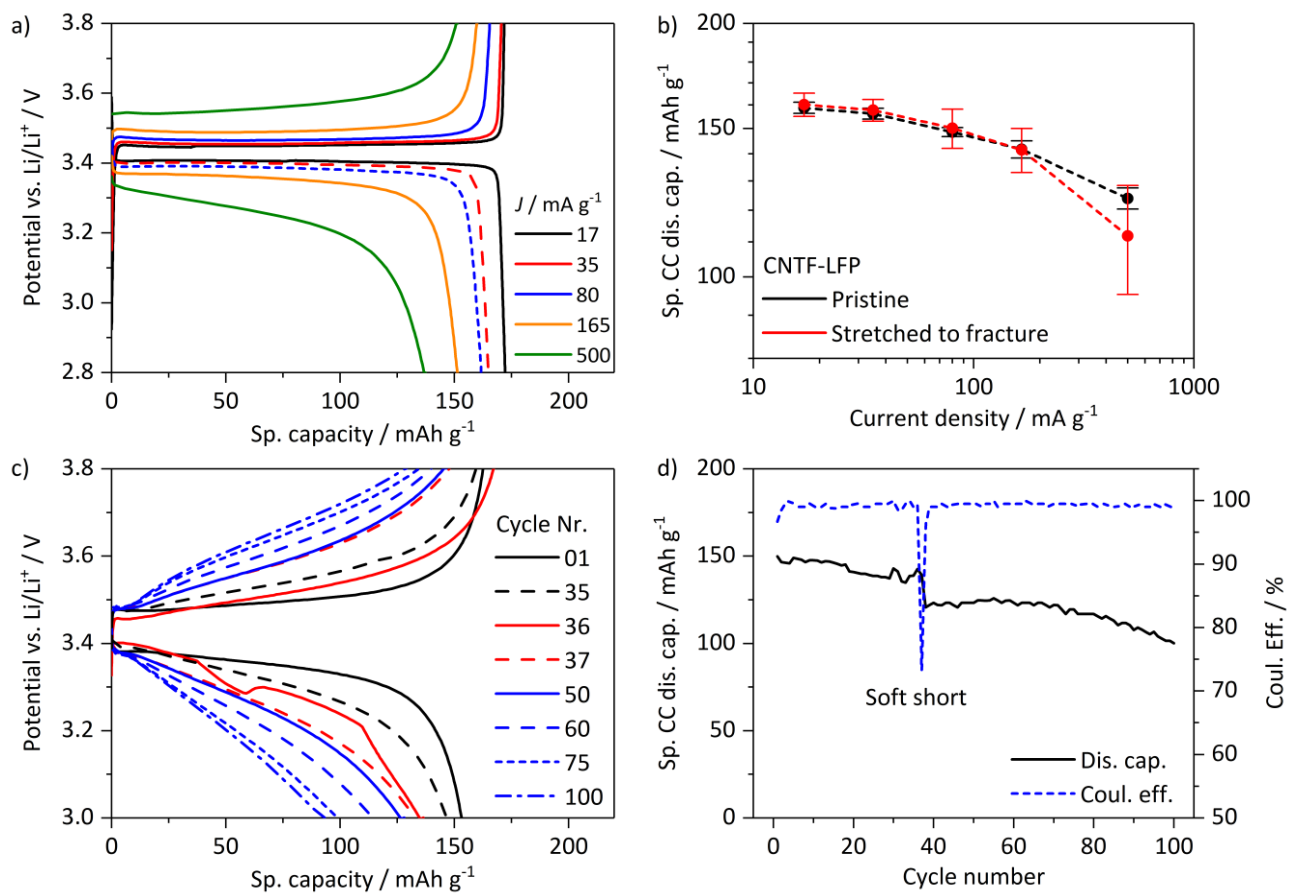

**Figure 6.** a) Potential profile of a CNTF-LFP/Li cell with fractured CNTF-LFP electrode, during the rate capability test; b) CC discharge capacity of pristine and stretched CNTF-LFP electrodes, at various current densities; c) Evolution of the cell potential profile of a CNTF-LFP/Li cell with fractured CNTF-LFP electrode during cyclability test at 80 mA g$^{-1}$. d) Specific discharge capacity (CC discharge till 3.0 V) and coulombic efficiency with a stretched CNTF-LFP electrode during the cyclability test.

Fitting of the capacity/rate curves of fractured CNTF-LFP electrodes with Eq. 2 (Figure S12) resulted in a worsening of the performance parameters, with time constant $\tau$ of 132 s and transport coefficient $\Theta$ of $6 \cdot 10^{-8}$ cm$^2$ s$^{-1}$ (Figure S12 and Table S2). Overall, these parameters are close to the ones obtained with Al-LFP electrodes. Interestingly, the exponential parameter $n$ is 0.58. This value is slightly higher than the one of pristine CNTF-LFP, but still lower than that of Al-LFP electrodes, which is close to 0.7. This indicates that predominant diffusive control is retained, despite the cracks and damages in the coating layer.

The cells with fractured electrodes show increasing polarization upon cycling (Figure 6c). Correspondently, the average capacity retention is lower than with pristine electrode, i.e. 70 % after 100 cycles, compared with 96 % for the undeformed electrode. The average coulombic efficiency is also slightly lower, decreasing from 99.5 % to 98.5 % after 100 cycles (Figure 6d).

One of the main sources of degradation of the cells with stretched electrodes is the occurrence of soft short circuits, observed as charge capacity spikes during cycling, as in Figure 6d at the 36th cycle. After the short, a drop in the CC discharge capacity is observed (compare cycles 35 and 36 in Figure 6c). These shorts are due to the inhomogeneous lithium deposition at the anode. Lithium dendrites develop faster with fractured CNTF-LFP cathodes due to the higher cell polarization. All six tested cells showed soft-shorting during cycling, and two cells tested failed irreversibly, after 50 and 70 cycles. The worsening of the cell cyclability is attributed to the damages introduced during stretching. The fractures observed by SEM cause progressive worsening of the electrical contact with the current collector. The higher cathode polarization causes in turn inhomogeneous plating at the anode and cycling instability.

Once again, it must be stressed that this problem arises from the extreme strain to which the tested electrodes have been subjected, which are not representative for ordinary operating conditions. Nonetheless, if necessary, this problem could be mitigated by implementing different strategies. For instance, the use of thicker CNTF current collector would result in considerably stronger electrode assembly, at a cost of a moderate decrease of the energy density. As an example, doubling the current collector thickness, while maintaining constant the active material loading, would theoretically result in a 90 % increase of the tensile strength, whereas the energy density would decrease by a mere 7 %. Alternatively, the use of elastic binders[55] or carbon nanotubes as conductive additive could result in a strengthening of the coating layer itself.[29,37] In a recent report,[56] Park et al. obtained extremely high areal capacities and thicknesses in Si and NMC electrodes, through the use of carbon nanotubes as both binder and conductive additive. The tailoring of the ratio between particle size and nanotube length resulted in the formation of segregated CNT networks, and in particularly favorable electrode morphology and mechanical properties. The use of CNTF current collectors could be combined with this last approach to strengthen the electrode coating and increase the areal capacity. However, it must be considered that the formation of segregated CNT networks requires the use of micrometer-sized active material particles. This could be detrimental if the active material has a low diffusion coefficient, as in the case of LFP ($10^{-16}$ - $10^{-15}$ $cm^2$ $s^{-1}$).[57]

Of the many multifunctional metrics, we propose to evaluate the combined mechanical and electrochemical properties of planar electrodes in terms of their (discharge) energy and tensile fracture energy, in both cases normalized by mass rather than volume and taking into account the current collector. Figure 7 presents relevant literature data for different active materials (CF, Si, LFP. LTO, LCO, graphite) and current collectors (CNTs, graphene and metals). The results presented in this work stand out because of the exceptional combination of energy density and toughness, together with acceptable cyclability even after fracture. But very importantly, the improvements reported here are directly transferable to a wide range of active materials, particularly those that also benefit from the high electrochemical stability of graphitic materials. In addition, the fabrication method used enables producing electrodes with different balance of electrochemical/mechanical performance, with properties governed by the rule of mixtures linear dependence shown in Figure 7.

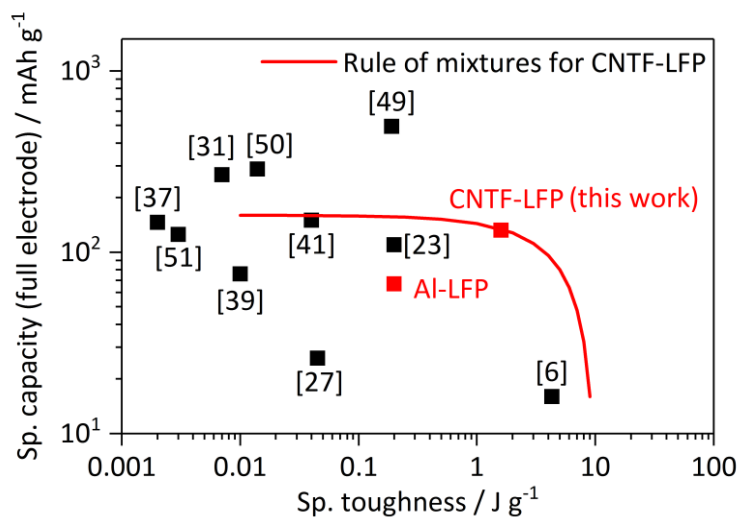

**Figure 7.** Comparison of specific capacity (normalized on the full-electrode-mass) and specific toughness (tensile fracture energy) of CNTF-LFP electrodes, Al-LFP electrodes, and literature data for flexible electrodes in planar configuration.

4. CONCLUSIONS

In this study, we reported on the electrochemical, mechanical and morphological properties of LFP electrodes with built-in CNTF current collector. The CNTF-LFP electrodes have high ductility and tensile strength, resulting in unprecedented electrode toughness, between one and two orders of magnitude above previous reports. Very importantly, these properties are combined with a > 90 % reduction of the current collector weight. The CNTF-LFP electrodes show excellent rate capability and cyclability, with discharge capacity of 124 mAh $g^{-1}$ at 500 mA $g^{-1}$ and 96 % capacity retention after 100 cycles at 80 mA $g^{-1}$. In the same cycling conditions, Al-LFP electrodes deliver 124 mAh $g^{-1}$, and show 85 % capacity retention.

These properties are a consequence of the partial infiltration of the active material into the porous CNTF current collector, forming essentially a composite structure with strong adhesion between elements and low electrical resistance at the interface. The improved electrical contact between current collector and active material allows an increase of the active material content which, together with the lower weight of the current collector, results in an enhancement of the energy density with respect to cells with Al-LFP electrodes with comparable mass loading. At a cell level, and at 500 mA g$^{-1}$, CNTF-LFP/Li cells deliver 78 mWh g$^{-1}$, versus 49 mWh g$^{-1}$ of the Al-LFP/Li cells.

Electrodes subjected to tensile tests up to fracture were extensively analyzed to determine their mechanical failure mode by SEM fractography, and their electrochemical performances by galvanostatic cycling. The tensile stress causes fractures in the coated layer, with cracks running transversal to the strain, and partial detachment of the coated layer from the current collector. The associated mechanism involves an initial uniform elastic deformation, followed by local fracture of the coating, at which point stress is transfer from the CNTF to the LFP coating in shear and the electrode strain is produced mainly at the CNTF with a non-uniform distribution. In spite of the large deformation experienced up to fracture, and of the formation of micro-cracks, the electrochemical properties are mostly retained. Stretched electrodes deliver 112 mAh g$^{-1}$ at 500 mA g$^{-1}$, still superior to undeformed Al-LFP electrodes.

The use of CNTF films as current collectors enables the development of electrodes with improved electrochemical and mechanical performances, crucial for application in flexible or stretchable batteries. In addition, the high toughness and ductility, and the retention of the electrochemical performances after tensile fracture, suggest the possibility of using these electrodes in stretchable batteries even in case of extreme stress and battery overstretching. Future work is directed at demonstrating the generality of the electrode architecture presented here, particularly with high energy density active materials and conducting similar studies in full cells.

SUPPORTING INFORMATION Digital images of the electrode processing, SEM images of in-house and commercial Al-LFP electrodes, resume graphs on the mechanical properties of Al-LFP and CNTF-LFP electrodes, stress-strain curve of free-standing LFP film, electrochemical performances of in-house and commercial Al-LFP electrodes, and of high-loading CNTF-LFP electrodes, fitting curves of capacity/rate curves, cyclability of CNTF-LFP and Al-LFP electrodes, impedance spectra, and calculation details of Table 2 entries.

ACKNOWLEDGEMENTS A.M. acknowledges funding from the European Union's Horizon 2020 research and innovation programme under the Marie Skłodowska-Curie grant agreement 797176 (ENERYARN). J.J.V. is grateful for generous financial support provided by the European Union Seventh Framework Program under grant agreement 678565 (ERC-STEM) and by the MINECO (RyC-2014-15115). N. B. acknowledges Comunidad de Madrid for the post-doctoral fellowship 2018-T2/AMB-12025. The authors thank Prof. C. González for discussions about the mechanics of these composite electrodes.

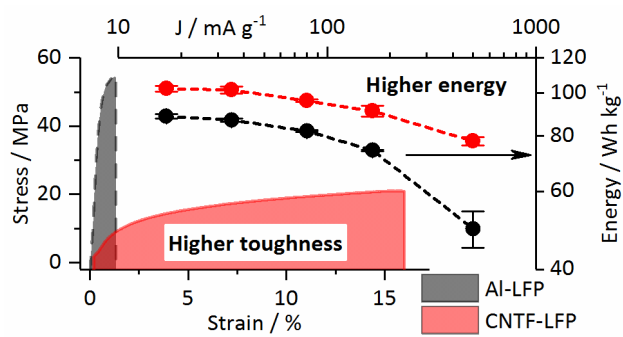